\documentclass[onecolumn,noshowpacs,nofootinbib,unicode,11pt]{revtex4}

\usepackage{amsfonts,amsmath,amssymb,tikz,accents}
\usepackage[eulergreek]{sansmath}
\usepackage{indentfirst,bigints}
\usepackage{feynmp-auto}
\usepackage{slashed}
\counterwithout{equation}{section}

\usepackage{textgreek}
\newcommand{\beq}{\begin{eqnarray}}
\newcommand{\eeq}{\end{eqnarray}}
\newcommand{\bea}{\begin{eqnarray}}
\newcommand{\eea}{\end{eqnarray}}
\newcommand{\p}{\partial}
\usepackage{slashed}

\begin{document}

\title{{ Renormalized Kalb-Ramond model: duality and generalized potential }}

\author{G.B. de Gracia}
\affiliation{Federal University of ABC, Center of Mathematics,  Santo Andr\'e, 09210-580, Brazil.}
\email{g.gracia@ufabc.edu.br}

\begin{abstract}
Considering the recent advances, the weak correlation between the massive Kalb-Ramond and the Proca interacting models is investigated by means of a set of complementary quantum field techniques beyond the semi-classical approach. A consistent framework to discuss the abrupt degree of freedom variation in the massless limit is established. In this manner, the Stueckelberg procedure is generalized for the present case to derive Ward-Takahashi constraints. By adding couplings with fermionic matter fields, even the weak dual correlation is broken due to the explicit form of the radiative corrections. However, considering the master action approach, the fermionic 1(PI) functions of the Proca electrodynamics are reproduced in all orders of perturbation.  A well-defined procedure to eliminate non-desirable modes emerging from quantum corrections is successfully applied for this model. Last, but not least, we prove that this structure can also provide a spin interaction which is relevant even at macroscopic distances, being a completion for a recently proposed confining potential. This is based on an interacting theory in compliance with renormalizability.
  
  {\bf{keywords: Duality, radiative corrections, massless limit,spin potential}}
\end{abstract}

\maketitle
\flushbottom

\section{Introduction }
\indent The so-called Kalb-Ramond model \cite{kalb} was originally studied as an excitation present in the string spectrum \cite{string}. Its massive phase describes spin $1$ particles, and the massless limit is associated to a scalar excitation. This abrupt degree of freedom jump is a consequence of the  reducible nature of the constraints \cite{redr,esc}  associated to the massless phase. \\
\indent The theory is described by a rank $2$ tensor field, leading to a wider set of possible couplings. It is suitable for applications in effective theories for low energy excitations of bound states with integer spin in QCD \cite{umr,tresr}. It is also of fundamental importance in the context of bosonization in $D=3+1$ dimensions \cite{boson} and discussions on magnetic monopole formulations \cite{mag}. Regarding condensed matter investigations, it leads to an alternative description \cite{fluid1} of superfluidity and quantum vortices. There are also recent developments associated with the comparison between the Gross–Pitaevsky description and the antisymmetric tensor field one \cite{fluid2}. In the context of gravity/cosmology, it is worth mentioning the  new achievements in the context of curved manifolds \cite{curved}, some exotic wormhole solutions \cite{worm} and the recent application in brane world scenarios \cite{brane}.\\
\indent Having in mind the indisputable applicability of the Kalb-Ramond model, we consider relevant to unveil some of its most intricate field theory features. It is well known that there is a kind of weak duality \cite{weak} between the massive phase and the Proca model \cite{proca}, as well as one between it and a scalar field for the massless limit \cite{spa}. The term weak is justified by the fact that although the same degrees of freedom are shared, the map between their propagators is obtained up to contact terms. The issue of duality has profound implications in physics, from condensed matter \cite{web,nastase} to AdS-CFT considerations \cite{ads}.  \\ 
\indent Recently, the reference \cite{god} revealed profound differences between the Proca and the massive Kalb-Ramond model in the interacting regime. Accordingly, the references \cite{god1,god2} investigated some correlated aspects.  However, there is still the possibility of discussing the intricate massless limit and also the development of a full quantum treatment beyond the semi-classical approach. To this end, we study this theory in an alternative structure, expressed in terms of a rank $2$ field, suitable to be generalized for $D$ dimensions \cite{umr,tresr,gab,barb,ddalm}. Although it is equivalent to the standard four-dimensional formulation \cite{spa,helay}, it allows a straightforward $D$ dimensional extension of this master action weak duality \cite{gab,m1,m2}. Accordingly, one can also mention another relevant recent work \cite{spa2} in which the Maxwell and  Kalb-Ramond models are conveniently coupled to reproduce, by means of a local Lagrangian, an effective potential for confinement in QCD.  The mass parameter plays an important role in the definition of the phase transition. This is an additional motivation to establish a well-defined background to study both massive and the massless phases in a unified description. Then, the possibility to extend this specific coupling to describe charge and also spin interaction potentials is another feasible goal developed in this work. The idea is to establish an interacting theory between Kalb-Ramond and Maxwell fields coupled with fermionic matter, whose associated low energy static potential displays these interactions.\\
\indent Therefore, our objective is to provide a set of complementary discussions to establish a clearer delimitation of these weak dual maps considering the effect of the radiative corrections as well as the non-perturbative constraints. The best way to test this duality is to include interactions and then search for non-compatible quantum corrections arising for both Proca and Kalb-Ramond models.  Considering the Proca model structure, we introduce the standard interaction with fermionic matter and a correlated derivative one for the Kalb-Ramond model according to master action suggestions. This kind of interaction is the most convenient to discuss these features.  We are also interested in defining the massless limit for the renormalized structure and how the degree of freedom jump is expressed in this context. \\
\indent In order to achieve these goals, an improved study of the massless limit is required. Moreover, we fully characterize the free system with a rigorous Heisenberg approach as a first step to introduce the standard path integral procedure to address the interacting case.\\
\indent More specifically, we intend to provide a Heisenberg quantization of the free system in order to covariantly define the transition between the massive/massless phase and explicitly display the jump in the physical modes. A crucial point is the fact that it is compatible with the well-known path integral formalism. Then, it can be understood as a complementary tool to establish a conceptual basis to investigate the radiative corrections associated to the interacting theory. These are obtained by means of standard field theory quantization schemes in the final sections of this paper.\\
\indent The Kugo-Ojima-Nakanishi (KON) indefinite metric formulation \cite{Nak1,Nak2} is convenient to be applied here since it leads to a simple covariant description of the target tensor model even with its associated non-trivial reducible constraint structure. Since the formalism consists in adding new auxiliary fields to turn the system into a second-class one, ensuring that the equations of motion and the constraints are valid in their strong forms, these complicated aspects can be consistently overcome. This is also an invitation to extend the formalism since two new auxiliary fields are required as well as a consistent way to get rid of them from the physical spectrum. This is achieved by defining a subsidiary condition capable to separate the positive norm subspace. As we are going to see, it also furnishes a suitable framework to obtain a smooth massless limit. One can also mention a variety of recent research associated with this formalism. It is a relevant tool with recent applications in the  investigation of quantum gravity in tetrad formalism\footnote{For which the standard approaches generally fail.} \cite{ref1}, the definition of a complementary tool for BRST symmetry extensions \cite{ref2},  for the discussion of mass generation mechanisms \cite{ref3}, in QCD confinement research \cite{conf} and to unveil formal aspects of QED$_4$ in the so-called non-linear t'Hooft gauge \cite{thooft}.\\
\indent Another necessary step is to introduce the generalization of the Stueckelberg procedure \cite{ddalm,stu,stu1,stu2} suitable for our $D$-dimensional formulation. A prescription to define a convergent path integral with all the required auxiliary fields is a correlated achievement. It provides a method to establish the theory in a degree of freedom conserving fashion to consistently describe the massless limit transition. The Stueckelberg procedure is also a useful artifice to discuss Ward-Takahashi constraints even for the massive case, leading to a relation for the mass renormalization constant and defining the general structure for the bosonic self-energy. Moreover, following the Gaussian map suggestions, a spinor interaction is introduced for both Kalb-Ramond and Proca fields, revealing deep differences in the qualitative as well as the quantitative features of their set of radiative corrections. Contrary to the first, the last model is non-renormalizable and has a quantum two-point structure that demands the inclusion of new counterterms to eliminate the divergent pieces. A consistent procedure to eliminate undesired poles induced by radiative corrections is developed here considering the concept of Merlin mode \cite{merlin}, and the approach of \cite{tresr} for a generalized counterterm Lagrangian. This framework is also a useful laboratory for applications in quantum gravity \cite{grav}.\\
\indent The Ward-Takahashi identity associated with the three-point function reveals not just a formal, but a deep physical difference between the interacting models. For a given non-linear generalization of the master action, although the bosonic sector is non-renormalizable, the purely fermionic 1(PI) functions are reproduced at all orders of perturbation theory. Finally, we take advantage of the present formulation to extend part of the seminal investigation of \cite{spa2} to derive a confining potential involving a spin interaction. To this end, we consider the recent developments of \cite{spind} in the ultra non-relativistic limit in a generalization of the coupling present in \cite{spa2}. Unlike the usual Pauli-like interaction, it leads to a static potential with no delta-like singularities, which is relevant even for macroscopic distances.  We prove that this theory can be embedded in an interacting Lagrangian in compliance with a set of renormalization principles. Moreover, as well as the confining structure, it is worth mentioning that the inclusion of this spin interaction is also motivated by the QCD phenomenology \cite{conf3,conf4,conf5,conf6,conf7,conf8}    \\
\indent The work is organized in the following manner. In Sec. \ref{sec2}, the (KON) formalism is generalized for the Kalb-Ramond model by the inclusion of two extra auxiliary fields. The definition of the physical subspace is carefully derived, as well as the complete set of commutators. In Sec. \ref{sec3}, the physical modes are covariantly separated and the massless limit and the weak duality constraints are discussed.  Later, in Sec. \ref{sec4}, the Stueckelberg procedure for this specific field content is introduced and generalized for $D$ dimensions.  In Sec. \ref{sec5}, a gauge-invariant interacting Lagrangian is considered and the Stueckelberg procedure is applied to derive Ward-Takahashi identities for the mass and wave function renormalization constants. The Sec. \ref{sec6} is devoted  to explicitly adding interactions in a well-defined way. Even the weak duality is broken by the radiative corrections, according to our analysis of the two-point sector and the self-energy tensor structure. A suitable treatment for the quantum corrections and the counterterms is provided in such a way that non-physical modes become unstable and do not contribute to the asymptotic spectrum. The massless limit for the physical pole is also investigated.
In Sec. \ref{sec7}, the discussion on the generalization of the master action approach demands a non-linear fermion self-interaction, suitably expressed by a Hubbard-Stratonovich transformation. In this way, all the Proca-like fermionic 1(PI) functions are reproduced by this dual model. 
In Sec. \ref{sec8}, we extend a previous proposal for a confining potential to also achieve spin interactions. The interacting theory is displayed in a structure compatible with a set of good quantum properties thanks to the presence of the auxiliary fields.
Finally, in Sec. \ref{sec9}, we conclude and present new perspectives. The metric signature (+, -, -, -) is used throughout. 

\section{Kalb-Ramond model quantization in (KON) framework }\label{sec2}

  \indent The (KON) formalism \cite{Nak1} is based on a Heisenberg quantization that includes extra auxiliary fields to turn the system into a second-class one. In this case, the constraints can be considered in their strong form and the correspondence principle can be consistently established without ambiguities. For the case of the Kalb-Ramond (KR) model \cite{kalb}, this procedure yields
  
  \bea {\cal{L}}=-\Big(\p^\mu \mathcal{B}_{\mu \nu}(x)\Big )^2+\frac{M^2}{2}\mathcal{B}_{\mu \nu}(x)\mathcal{B}^{\mu \nu}(x)+\varepsilon^{\mu \nu \alpha \beta}\phi_\mu(x) \p_\nu \mathcal{B}_{\alpha \beta}(x)+\p^\mu \phi_\mu(x)b(x) \eea
In this work, the auxiliary vector, the scalar fields, and also the gauge parameters are pseudo-vector or pseudo-scalar fields while the rank $2$ boson field indeed transforms as a tensor. This is required to achieve a Lagrangian with standard scalar nature. The present formulation \cite{umr,tresr} can be associated to the standard Kalb-Ramond \cite{kalb} structure as $\epsilon^{\mu \nu \alpha \beta}\mathcal{B}_{\alpha \beta}(x)=H^{\mu \nu}(x)$ for $D= 3+1$ dimensions.  However, just the present formulation can be straightforwardly generalized for $D$ dimensions keeping the rank $2$ structure with the same particle content. \\
  \indent A pertinent observation is that the classical Lagrangian has a massless phase which is invariant under the local transformation
  \bea \mathcal{B}_{\mu \nu}(x)\to \mathcal{B}_{\mu \nu}(x)+\varepsilon_{\mu \nu \beta \gamma}\p^\beta \Lambda^\gamma(x)\eea
  defining the redundant non-physical sector of the theory  for both massless and massive cases.\\
   \indent The (KON) formalism is a useful guide to unveil the physical degrees of freedom associated with the highly non-trivial kinetic operator associated with the Kalb-Ramond model. The massless phase has reducible constraints associated with the existence of zero modes that also have their own set of null modes. Therefore, since this quantization method depends on the introduction of extra auxiliary fields to turn the system into a second-class one, it provides a suitable framework to investigate the massless limit. This is due to the fact that both the massive and the massless models share the same derivative structure, implying the necessity of the same set of Lagrange multiplier fields. These are the ones that can immediately define the gauge conditions for the massless phase. Since it is based on an indefinite metric Hilbert space, it is always possible to define subsidiary conditions to eliminate the auxiliary fields from the observable spectrum, as we are going to show.\\
   \indent The canonical momenta are the following,
  \begin{align} \pi_{0i}(x)&=-2\Big(\p_0\mathcal{B}_{0i}(x)+\p^j\mathcal{B}_{ji}(x)\Big),\quad
   \pi_{ij}(x)=\varepsilon_{0ijk}\phi^k(x) \\
   \pi_\mu^{(\phi)}(x)&=\delta_0^\mu b(x), \quad 
   \qquad \qquad \qquad \qquad \pi^{(b)}(x)=0\label{ar}\end{align}
   providing a second-class structure and proving that the two extra fields were indeed necessary to achieve this goal.\\
   \indent In higher dimensions, a wider quantity of auxiliary fields is demanded. Therefore, for completeness, in $D\geq 2+1$ dimensions, the auxiliary part of the action is expressed as
\begin{align}
   S_{aux.}^D= \int \Big[\big(d\mathcal{B}(x)\big)\wedge \big(*\Phi^{(D-3)}\big)+\sum_{n=1}^{(D-3)}\big(\Phi^n\big)\wedge \big(*d\Phi^{(n-1)}\big)\Big] \end{align}
   in which $\Phi^n(x)$ denotes a $n$-form auxiliary field with an appropriate discrete symmetry parity to ensure a scalar action. The gauge field $2$-form is defined as $\mathcal{B}\equiv \frac{1}{2}\mathcal{B}_{\mu \nu}dx^\mu \wedge dx^\nu$ \\
   \indent The complete set of extra fields, collectively denoted as $\Phi$ imply a second-class Lagrangian suitable for the consideration of the correspondence principle. In the functional quantization language, it is associated with a well-established measure for a convergent path integral. The $D=3+1$ dimensional theory is immediately recovered from the general expression. The symbol $*$ represents the Hodge dual star operation.\\
   \indent Returning to the $D=3+1$ dimensional case, the Dirac brackets defined on the reduced phase space are associated with the primary constraint matrix
   \bea \mathcal{G}^{IJ}(x,y)=\{\mathcal{C}^I(x),\mathcal{C}^J(y)\}=\left(\begin{array}{ccccc}
	0& -1& 0& 0      \\
	1 & 0 & 0 & 0       \\
	0& 0& 0   &-\varepsilon_{ijn}    \\
        0& 0& \varepsilon_{xym}    &0     \\
\end{array}\right)\delta^3(x-y)  \eea
with constraints conveniently grouped as
\bea \mathcal{C}^I(x)=\left(\begin{array}{cccccccc}
	\pi^{(b)}(x)& \\
	\pi_0^\phi(x)-b &     \\
	\pi_{mn}(x)-\varepsilon_{mnl}\phi^l(x)  &      \\ 
         \pi_n^\phi(x) &     \\
\end{array}\right)\eea
associated with the Dirac brackets 

\bea \{F(x),G(y)\}_D=\{F(x),G(y)\}-\int d^3w d^3z\{F(x),\mathcal{C}^I(w)   \}\tilde{\mathcal{G}}_{IJ}\{\mathcal{C}^J(z),G(y)   \}\eea
with $F(x)$ and $G(x)$ being arbitrary phase space functionals.\\
\indent After this digression, returning to the main goal of this section, the reduced brackets explicitly read
 \begin{align}\Big\{ \mathcal{B}_{ij}(x),\varepsilon_{0mnk}\phi^k(y)     \Big \}&=\Delta_{ij}^{\ mn }\delta^3(x-y),\quad
  \Big\{ \mathcal{B}_{0i}(x),\pi^{0k}(y)     \Big \}=\delta_i^k\delta^3(x-y)\\
   \Big\{ \phi^i(x),\pi^{(\phi)}_j(y)     \Big \}&=\delta_j^i \delta^3(x-y),\quad \qquad \quad
   \Big\{ \phi_0(x),b(y)     \Big \}= \delta^3(x-y)
   \end{align}
   denoted without the $D$ subscript for simplicity. They are adequate to proceed with the correspondence principle. Here, $\Delta_{ij}^{mn}\equiv \frac{(\delta_i^m \delta_j^n-\delta_i^n \delta_j^m )}{2} $.\\
    \indent Owing to the correspondence principle, the initial conditions for the commutators
  \begin{align}  \Big[\mathcal{B}_{0l}(x),\p_0\mathcal{B}^{0i}(y)\Big]_0&=-\frac{i}{2}\delta_l^i\delta^3(x-y),\\
  \Big[\mathcal{B}^{kl}(x),\phi^m(y) \Big]_0&=-\frac{i}{2}\varepsilon^{klm}\delta^3(x-y),\ \quad 
  \Big[\phi_0(x),b(y)\Big]_0=i\delta^3(x-y)\end{align}
  can be obtained from the Dirac brackets algebra.\\
  \indent The operator equations of motion are
  
  \begin{align} -\varepsilon^{\mu \nu \alpha \beta}\p_\nu &\phi_\mu(x)+\Big[ \p^\alpha(\p_\nu \mathcal{B}^{\nu \beta}(x))-\p^\beta(\p_\nu \mathcal{B}^{\nu \alpha}(x))\Big] +m^2\mathcal{B}^{\alpha \beta}(x)=0\label{eq1}\\
  &\p^\mu \phi_\mu(x)=0\qquad  \qquad
  \varepsilon^{\mu \nu \alpha \beta}\p_\nu \mathcal{B}_{\alpha \beta}(x)-\p^\mu b(x)=0\label{eq2} \\ \nonumber
\end{align}
\indent These equations can be applied to derive the following extra initial conditions
\begin{align} &\Big[\p_0\phi_l(x), \phi^k(y)\Big]_0=-\frac{i}{2}M^2\delta_k^l\delta^3(x-y)\nonumber \\
&\Big[\p_0\phi_l(x), \phi^k(y)\Big]_0=-\frac{i}{2}M^2\delta_k^l\delta^3(x-y)\end{align}

 \indent The divergence of the eq.(\ref{eq1}) leads to the propagating $s=1$ massive mode
   \bea \Big(\Box+M^2\Big)\p^\nu \mathcal{B}_{\nu \mu}(x)=0\eea
   \indent The divergence of the last equation of (\ref{eq2})  furnishes a wave equation for the scalar auxiliary field
   \bea \Box b(x)=0\eea
   \indent Then, one can also derive the equation
   \bea \Box^2 \phi_\mu(x)=0\eea
 considering the previous relations.\\
    \indent Taking into account the operator equation due to the auxiliary vector field, yields
  \bea \Big[\p_0b(x), b(y)\Big]_0=0\eea
  \indent Regarding the commutator at unequal times, it should obey
  \bea \Box^x \Big[b(x), b(y)\Big]=0\eea
  \indent In order to solve this equation, we introduce the following integral representation
 \begin{align}
 F(x-y)&= \int d^4u \ \varepsilon(x,y,u) \tau(x-u)G(u-y) \nonumber \\  
 \quad-&\int d^3u \Big[ \tau(x-u)\p_0^uF(u-y)-\p_0^u\tau(x-u)F(u-y) \Big]_{u^0=y^0}
 \label{ghostTwoPointr}
 \end{align}
 in which  $\varepsilon(x,y,u)$ is defined as
 \begin{equation}
 \varepsilon(x,y,u)= \Theta(x_0-u_0)- \Theta(y_0-u_0).
 \end{equation}
 for an operator obeying the relation $\hat{O}^x F(x-y)=G(x-y)$ such that $\hat{O}^x\tau(x-y)=0$, with $\hat{O}^x$ being a given differential operator and $\tau(x-y)$  its Green function.\\
 \indent Therefore, it leads to the zero norm condition
 \bea  \Big[b(x),b(y)\Big]=0      \eea
\indent Accordingly, the following set of commutators can be derived considering an analogous approach
\bea \Big[\phi_\mu(x),\phi_\nu(y)\Big]=\frac{i}{2}M^2\Big(\eta_{\mu \nu}\Delta(x-y,0)-\p_\mu \p_\nu E(x-y,0)\Big)\eea

\bea \Big[\phi_\mu(x),b(y)\Big]=-i\p_\mu \Delta(x-y,0),\quad \quad \Big[\mathcal{B}_{\mu \nu}(x),\phi_\alpha(y)\Big]=\frac{i}{2}\varepsilon_{\mu \nu \alpha \beta}\p^\beta \Delta(x-y,0)\eea

 \begin{align} \Big[\mathcal{B}_{\mu \nu}(x),\mathcal{B}^{\alpha \lambda}(y)\Big]=&-\frac{i}{2M^2}\Big(\eta_\mu^\alpha\p_\nu \p^\lambda+\eta_\nu^\lambda\p_\mu \p^\alpha-\eta_\mu^\lambda\p_\nu \p^\alpha-\eta_\nu^\alpha \p_\mu \p^\lambda\Big)\Big(\Delta(x-y,M^2)-\Delta(x-y,0)\Big)\nonumber \\ \label{comut}\end{align}
with the Pauli-Jordan distribution $\Delta(x-y,M^2)$  and the double pole one $E(x-y,M^2)$ defined in the appendix A.\\
 \indent Considering this commutator structure, the following subsidiary condition avoids observing non-physical  projections of the antisymmetric field and also eliminates the presence of the auxiliary fields in the positive semi-definite subspace 
  \begin{equation}
  \phi_\mu^+(x) | \text{phys} \rangle = 0, \quad \forall | \text{phys} \rangle \in \mathcal{V}_{\text{phys}}.
  \label{conditionBfreefieldr}
  \end{equation}
  in which $\phi_\mu^+(x) $ denotes the positive frequency part of the vector auxiliary field. Since the states do not have space-time dependence in the Heisenberg description, the definition above is Poincaré invariant.\\
  \indent The observable physical sector is defined up to zero norm states contained in $\mathcal{V}_0$
  \bea \mathcal{H}_{\text{phys}}=\frac{\mathcal{V}_{\text{phys}}}{\mathcal{V}_0} \eea
 \indent Then, one can easily conclude that $\p_\mu \mathcal{B}^{\mu \nu}(x)$ is contained the physical subspace, being an observable. The remaining auxiliary fields are not physical, as they should be.\\
 \indent Regarding the massless limit, there is the following expansion  $\Delta(x-y,M^2)=\Delta(x-y,0)-M^2E(x-y,0)+...$. It is a useful identity to establish the massless limit and investigate the jump in the degrees of freedom.
  
  \section{On the observable pole fields and massless limit discussion}\label{sec3}

  \indent The (KON) formalism provides a useful framework to discuss the massive/massless phase transition. The effect of the auxiliary fields implies well-defined commutators, even in the massless limit. Then, it is instructive to analyze the commutator between the massive pole fields
  
  \bea \Big[\p^\mu \mathcal{B}_{\mu \nu}(x),\p^\alpha \mathcal{B}_{\alpha \lambda}(y)\Big]=-\frac{i}{2M^2}\Big(M^4\eta_{\nu \lambda}+M^2\p_\nu \p_\lambda\Big)\Delta(x-y,M^2)\eea
  \indent This commutator indicates a spin-$1$ content. It is also possible to notice that the massless phase has a longitudinal scalar nature
  \bea\Big[\p^\mu \mathcal{B}_{\mu \nu}(x),\p^\alpha \mathcal{B}_{\alpha \lambda}(y)\Big]=-\frac{i}{2}\p_\nu \p_\lambda \Delta(x-y,0)\eea
  \indent This is in agreement with the analysis of \cite{gab} in which the transition from the spin-$1$ to the spin-$0$ regime in the massless limit was investigated by phase space degree of freedom counting and also the analysis of the residue of the saturated amplitude. The pole  move from the spin-$1$ to the spin-$0$ sector at the mentioned limit, representing a variation of two degrees of freedom.\\
  \indent Regarding the nature of the projections of the vector pole in the massive/massless phase, it is useful to explicitly consider their Fourier component algebra associated with its positive frequency part. The massive phase yields
  \bea      \Big[ \tilde a_{\mu}(p), \tilde a_{\nu}^\dagger(q)        \Big]= \frac{1}{2}\theta(p_0)\big(-M^2\eta_{\mu \nu}+p_\mu p_\nu   \big)\delta(p^2-M^2)\delta^4(p-q)      \eea
\indent Assuming the frame $p_\mu=\big(M,0,0,0 \big)$, one concludes that the three spatial components have positive norm and the temporal one has null projection. This is compatible with the spin-$1$ content of this phase. Its commutator structure is in agreement with the fact that there is a master action weak correspondence between the divergence of the tensor field and the Proca one \cite{spa}
\bea A_\mu \longleftrightarrow \frac{\sqrt{2}}{M}\p^\lambda \mathcal{B}_{\lambda \mu}\eea
\indent The massless limit yields the phase
\bea      \Big[ \tilde a_{\mu}(p), \tilde a_{\nu}^\dagger(q)        \Big]= \frac{1}{2}\theta(p_0)p_\mu p_\nu \delta(p^2)\delta^4(p-q)     \eea
in accordance with the weak dual map derived in the master action approach 
\bea \p_\mu \Phi \longleftrightarrow \p^\lambda \mathcal{B}_{\lambda \mu}\eea
associating the tensor field with a scalar excitation.\\
\indent Considering the frame $p_\mu=\big(p_3,0,0,p_3\big)$, both the zeroth and the third component have positive projections, while the remaining components have zero norms. Due to the transverse constraint on this observable pole field, these components are linearly dependent. Therefore, there is just one degree of freedom in accordance with the expected spin-$0$ nature of the massless phase.

\section{On the generalized Stueckelberg procedure}\label{sec4}

\indent In the reference \cite{spa}, the Stueckelberg procedure is successfully established for the Kalb-Ramond field. However, it cannot be straightforwardly generalized for $D$ dimensions without altering the couplings and the tensor structure  of the physical fields. Otherwise, the particle content is violated. In \cite{ddalm}, an interesting proposal for a $D$ dimensional prescription is provided by means of a dimensional reduction approach. In order to achieve the full quantum version, we derive its associated set of auxiliary fields to define a convergent path integral. It can be established by means of the (KON) principle, based on the requirement to define a second-class system in a covariant manner. \\
\indent The present formulation for the Stueckelberg procedure can be established as
\begin{align}
 &S_{Stueck.+aux.}^D=\int \Bigg[ \frac{M^2}{2} \Big(\mathcal{B}-\frac{1}{M}\big(*d A^{(D-3)}\big)\Big) \wedge *\Big(\mathcal{B}-\frac{1}{M}\big(*d A^{(D-3)}\big)\Big)   +\big(A^{(D-3)}\big)\wedge *\big(d\chi^{(D-4)}\big) \nonumber \\&+\sum_{k=1}^{(D-4)} \chi^k \wedge*\big(d \chi^{(k-1)}\big)      +\Big[\big(d\mathcal{B}(x)\big)\wedge \big(*\Phi^{(D-3)}\big)+\sum_{n=1}^{(D-3)}\big(\Phi^n\big)\wedge \big(*d\Phi^{(n-1)}\big)\Big]         \Bigg]\end{align}
in terms of $p$-forms for simplicity. Here, $\Phi$ and $\chi$ forms denote auxiliary Lagrange multiplier fields. The complete set of auxiliary fields is displayed, the (KON) ones and also the specific Lagrange multipliers for the Stueckelberg procedure.\\
\indent The semi-classical theory without the Lagrange multiplier terms is associated with the following local freedom
\bea \delta A^{(D-3)}=d \sigma^{D-4}+M\Lambda^{(D-3)},\quad \delta \mathcal{B}=*\big( 
  d\Lambda^{(D-3)}\big)\eea
  \indent This formulation allows the obtainment of a convergent path integral with a finite number of auxiliary fields for any dimension $D\geq 2+1$. The massless limit is well-defined. \\
\indent In $D=3+1$ dimensions, the generalized Stueckelberg procedure \cite{stu1,stu2} can be established for this particular model by the cost of adding a pseudo-vector auxiliary field
\begin{align} S= \int d^4x\Big[& -\Big(\p^\mu \mathcal{B}_{\mu \nu}\Big )^2+\frac{M^2}{2}\Big(\mathcal{B}_{\mu \nu}+\frac{1}{M}\varepsilon_{\mu\nu \alpha \beta}F^{\alpha \beta}\Big)^2 \\ \nonumber & \varepsilon^{\mu \nu \alpha \beta}\phi_\mu(x) \p_\nu \mathcal{B}_{\alpha \beta}(x)+\p^\mu \phi_\mu(x)b(x) +\p_\mu A^\mu B
 \Big] \end{align}
with  $F_{\mu \nu}(x)\equiv \p_\mu A_\nu(x)-\p_\nu A_\mu(x)$.\\
\indent The pure gauge field transforms as \footnote{with $\Lambda_\mu(x)$ and $\sigma(x)$ being arbitrary fields.}
\bea  A_\mu(x)\to A_\mu-M\Lambda_\mu(x)+\p_\mu \sigma(x) \eea

in order to ensure an invariant Lagrangian  under the local transformation
\bea \delta \mathcal{B}_{\mu \nu}(x)=\varepsilon_{\mu \nu \alpha \beta}\p^\alpha \Lambda^\beta(x)\eea
in the absence of the auxiliary fields. If they are present, there is still a set of residual symmetries.\\
\indent Since $A_\mu(x)$ is invariant under two local transformations, the auxiliary field $B$ must be added to define a convergent path integral. Equivalently, in the (KON) formalism framework, it is included as a requirement to define a second-class system.\\
\indent It is well known that the Lagrangian above propagates three degrees of freedom \cite{gab} and describes a spin-$1$ particle. The auxiliary fields are non-observable.
However, taking the massless limit, $A_\mu(x)$ is not a pure gauge field anymore, and the following action is obtained

\begin{align} S= \int d^4x\Big[ -\Big(\p^\mu \mathcal{B}_{\mu \nu}\Big )^2-2F_{\mu \nu} F^{\mu \nu}+\varepsilon^{\mu \nu \alpha \beta}\phi_\mu(x) \p_\nu \mathcal{B}_{\alpha \beta}(x)+\p^\mu \phi_\mu(x)b(x)+\p_\mu A^\mu B \Big] \end{align}

\indent Up to auxiliary fields, the first term represents a scalar excitation, while the second one propagates two degrees of freedom. Therefore, the generalized Stueckelberg procedure allied with (KON) formalism indeed generates
a degree of freedom conserving structure suitable to investigate the transition to the massless limit.

\section{Stueckelberg Procedure as a Method to Derive Ward-Takahashi Identities}\label{sec5}

\indent The Stueckelberg procedure is a useful artifice to derive Ward-Takahashi identities between renormalization constants for massive models, since it creates an artificial gauge symmetry. Therefore, this is the ideal setting to include both massive and massless phases in the discussion.\\
\indent For now on, we consider an interaction Lagrangian $\mathcal{L}_I$ invariant under gauge transformations, parameterized by $\Lambda_\mu(x)$. This, and the presence of the Stueckelberg field, allows the obtainment of 
non-perturbative relations for the renormalization constants via Ward-Takahashi identities. After adding c-number sources for the fields, the functional generator becomes
\begin{align}   Z&=\int  \mathcal{D}\mathcal{B}_{\mu \nu} \mathcal{D}\mu_I{\cal{D}}A_\nu\mathcal{D}B{\cal{D\phi_\mu}}{\cal{D}}b \exp\Big[i\int d^4x\Big( -\Big(\p^\mu \mathcal{B}_{\mu \nu}\Big )^2+\frac{M^2}{2}\Big(\mathcal{B}_{\mu \nu}+\frac{1}{M}\varepsilon_{\mu\nu \alpha \beta}F^{\alpha \beta}\Big)^2\nonumber \\&+\varepsilon^{\mu \nu \alpha \beta}\phi_\mu \p_\nu \mathcal{B}_{\alpha \beta}  +\p^\mu \phi_\mu b+\p_\mu A^\mu B+\mathcal{L}_I+A_\mu J^\mu+\mathcal{B}_{\mu \nu}T^{\mu \nu}+...\Big) \Big]  \label{eqboa}       \end{align}
with ellipsis denoting the kinetic operators as well as source terms for the matter fields present in $\mathcal{L}_I$ that are not important for the next discussion. Here, $\mathcal{D}\mu_I$ denotes the measure associated to these matter fields. It is important to mention that in this model, as well as in the Proca one, there is no local $U(1)$ symmetry associated to the standard minimal coupling prescription.    \\
\indent From the invariance of the functional generator under this gauge transformation, it is possible to derive the following Ward-Takahashi identity
\bea \varepsilon^{\mu \nu \sigma \gamma}\p_\sigma^x\frac{\delta^2\Gamma}{\delta \mathcal{B}_{\mu \nu}(x)\delta \mathcal{B}_{\alpha \beta}(y)}=M\frac{\delta^2\Gamma}{\delta A_\gamma(x)\delta \mathcal{B}_{\alpha \beta}(y) } \eea
with $\Gamma$ representing the effective action.\\
\indent It is a coupled relation. Then, the knowledge of the Stueckelberg field quantum equations 
\bea \frac{\delta^2\Gamma}{\delta A_\gamma(x)\delta \mathcal{B}_{\alpha \beta}(y) } =M\varepsilon^{\gamma \nu \alpha \beta}\p_\nu \delta^4(x-y) \eea
is demanded to derive an identity for the tensor field inverse propagator.\\
\indent This identity implies the following form for the full bosonic self-energy
\bea \Pi_{\mu \nu , \alpha \beta}(p)=P^{[1e]}_{\ \mu \nu, \alpha \beta}\ \sigma(p)  \eea
in momentum space, with $\sigma(p)$ being a given scalar function. The projector $P^{[1e]}_{\ \mu \nu, \alpha \beta}$, associated to the physical sector, is defined in appendix B. \\
\indent Then, collectively denoting renormalized fields as $\Phi_A(x)=\sqrt{Z}_A\Phi_A^R(x)$ and the mass parameter as $m^2=Z_m m_R^2$, implies that $Z_mZ_\mathcal{B}=1$. One can also conclude $Z_A=1$, in accordance with the fact that the Stueckelberg field does not interact, being just an artifice.\\
\indent The massless limit can be smoothly obtained, preserving the degrees of freedom. Moreover, the presence of the (KON) auxiliary fields guarantees a convergent measure. The $A_\mu(x)$ field decouples and becomes a physical independent excitation. Indeed, one can also verify that no quantum correlation between this and the tensor field remains even considering the full radiative corrections. 

\section{Weak duality breaking by quantum corrections}\label{sec6}

\indent A possible way to test the weak duality between the Kalb-Ramond and Proca models is to consider a specific form for the interactions and then analyze the structure of the radiative corrections. Since Proca Electrodynamics is well known, one can compare it with the Kalb-Ramond version. Then, the interacting model for the Kalb-Ramond electrodynamics is the following
\begin{align}   Z&=\int \exp\Big[i\int d^4x\Big( -\Big(\p^\mu \mathcal{B}_{\mu \nu}\Big )^2+\frac{M^2}{2}\mathcal{B}_{\mu \nu}\mathcal{B}^{\mu \nu}+\varepsilon^{\mu \nu \alpha \beta}\phi_\mu \p_\nu \mathcal{B}_{\alpha \beta}\nonumber \\&  +\p^\mu \phi_\mu b+\frac{1}{m}\p^\mu\mathcal{B}_{\mu \nu}\big(e\bar{\Psi} \gamma^\nu \Psi\big)+\Bar{\Psi}\Big(i\slashed{\partial}-m_e\Big) \Psi  \Big) \Big] \mathcal{D}\Psi \mathcal{D}\Bar \Psi  \mathcal{D}\mathcal{B}_{\mu \nu}{\cal{D}\phi_\mu}{\cal{D}}b       \label{um}\end{align}

derived after the addition of a spinor non-linear coupling and the standard kinetic part for the fermions. $m$ is a parameter with mass dimension and $e$ denotes the electric charge. Although useful for the specific goal of discussing Ward-Takahashi identities, the Stueckelberg procedure is not considered in this present discussion, since its associated fields can be completely integrated out. \\
\indent Since  the vertex diagram has the same topology as the vector QED$_4$, the relations \footnote{V denotes the number of vertexes, $P_\gamma $ is the number of bosonic propagators, $P_e$ the number of fermionic propagators, with $N_\gamma$ and $N_e$ being the number of bosonic and fermionic external lines, respectively. $L$ denotes the number of loops.} $V=2P_\gamma +N_\gamma=P_e+\frac{1}{2}N_e$ and $L=P_e+P_\gamma-V+1$ are kept. The expression for the superficial degree of divergence becomes $D=4L+\big(V-N_\gamma \big)-P_e-2P_\gamma$, the coefficient $2$ for the boson internal line is due to the second order structure of the Kalb-Ramond propagator
\bea \mathcal{P}_{\mu \nu \alpha \lambda}^0(p)=-i\frac{P^{[1e]}_{\mu\nu,\alpha\lambda}}{\big(p^2-M^2\big)}\label{prop}\eea 
in terms of the spin projectors defined in the appendix B.\\
\indent This propagator structure is ensured by the effect of the auxiliary (KON) fields. It is indeed the Feynman version of (\ref{comut}), as it should be. One can notice that no singular behavior occurs in the massless limit, unlike the conventional treatment for massive gauge fields. The entire structure is proportional to the $P^{[1e]}_{\mu\nu,\alpha\lambda}$ projector in which the physical poles appear.\\
\indent The term  $\big(V-N_\gamma \big) $ expresses the fact that since the momentum is attached to the Kalb-Ramond field line when it is internal, the vertex leads to an increase in the divergence order of the graph. On the other hand, for the case in which it is external, there is no contribution to the divergence at all. Therefore, these constraints lead to $D=4+V-2N_\gamma-\frac{3}{2}N_e$. This dependence on the number of vertices implies a non-renormalizable nature for the interacting system, in opposition to the case of the Proca theory coupled to fermions. Therefore, this duality is explicitly violated by quantum corrections.\\
\indent Regarding the previous section,  one can consider Ward-Takahashi identities \footnote{For the case of the theory in the Stueckelberg framework.} to derive the following constraint on the vertex function
\bea \epsilon^{\beta \alpha\sigma \gamma}k_\sigma \frac{\delta^3\Gamma}{\delta \bar \Psi(p)\delta \Psi(\tilde p)\delta \mathcal{B}_{\beta \alpha}(k)}=0 \eea
\indent This is another result in contradiction with the interacting Proca model. The relation above do not force an exclusive constraint between the charge and photon renormalization constants, unlike the standard QED-like models. Considering these cases, this constraint implies that the proton and the electron have exactly the opposite charge, despite the fact that they participate in a set of completely different interactions. However, for the interacting Kalb-Ramond model, the charge renormalization is not necessarily dictated by the gauge field one. 

\subsection{Renormalization and asymptotic modes}

\indent Although not equivalent to Proca theory, this interacting model has interesting properties by itself. It can still be useful for an effective theory valid up to a given cutoff $\approx \frac{m}{e}$.  Since it is possible to derive a relation between the Kalb-Ramond model and the description of a quantum perfect fluid \cite{fluid2}, the quantum corrected system can be applied to describe a renormalized fluid associated with a Kalb-Ramond plasma.\\
\indent In order to establish the on-shell conditions for a non-renormalizable theory, a set of counterterms whose structure differs from the bare part is demanded. Here, they are suitably chosen to eliminate non-physical extra poles emerging due to the radiative corrections \cite{tresr} and to guarantee the unit residue for the physical excitations. The presence of interactions allows this cancelation of non-desirable modes in the so-called Merlin mode framework \cite{merlin}. The present procedure can be useful for a wide set of theories, including quantum gravity \cite{grav}.\\
\indent The bosonic quadratic part of the quantum action reads

\begin{align} \Gamma[\mathcal{B}]=\int \frac{d^4p}{(2\pi)^4}\Bigg\{\ \frac{\mathcal{B}^{\mu \nu\ }(p)}{2}\Bigg[   P^{[1b]}M^2-P^{[1e]}\Big(p^2-M^2+\frac{p^4}{m^2}\tilde \pi(p^2)\Big)\Bigg]_{\mu \nu \alpha \beta}\mathcal{B}^{\alpha \beta \ }(-p)\Bigg\}\end{align}
up to the auxiliary field sector.\\
\indent This result is obtained considering the following structure $\Pi_{\mu \nu}(p)=-p^2\theta_{\mu \nu}\tilde \pi(p^2)$ for the boson self-energy, with the transverse projector $\theta_{\mu \nu}(p)$ defined in the appendix B.\\
\indent By means of dimensional regularization, it is possible to derive the one-loop approximation \cite{sch}
\bea  \tilde \pi(p^2)=\lim_{\epsilon \to 0}\frac{e^2}{2\pi^2}\int dx  \big(1-x\big)x\Big[ \frac{2}{\epsilon}+\ln\Big(\frac{4\pi e^{-\gamma_E}\mu^2}{m^2_e-p^2x(1-x)}   \Big) \Big] \eea
with $\mu$ representing an energy scale and $\gamma_E$ being the Euler-Mascheroni constant. The parameter $\epsilon$ is defined as a small shift on the space-time dimensionality  $D=4-\epsilon$.\\
\indent Renormalizing the system imply in the following set counterterms
\bea \Gamma_{C.T.}=  \int \frac{d^4p}{(2\pi)^4}\Bigg\{\ \frac{\mathcal{B}^{\mu \nu\ }(p)}{2}\Bigg[   -P^{[1e]}\Big(p^2\delta Z_{\mathcal{B}}-\frac{p^4}{m^2}(1+\tilde \pi^{div.}(p^2))\Big)\Bigg]_{\mu \nu \alpha \beta}\mathcal{B}^{\alpha \beta \ }(-p)\Bigg\}
                \eea
                involving the divergent piece of the polarization tensor. Extra higher derivative terms are also suitably included in order to eliminate divergent structures and to keep the asymptotic spectrum.  The counterterms from $\Gamma_{C.T.}$ are in accordance with the previously derived Ward-Takahashi identities. The label $R$ for the renormalized structures is omitted for simplicity of notation. \\
\indent Then, for the renormalized system $\Gamma^R[\mathcal{B}]=\Gamma[\mathcal{B}]+\Gamma_{C.T.}$, the momentum dependent poles are defined as
\bea             m^2_{\pm}= \frac{m^2}{2(1-\tilde{\pi}^{fin.}(p))}\Bigg[(1+\delta Z_\mathcal{B})\mp \sqrt{ (1+\delta Z_\mathcal{B})^2-\frac{4M^2}{m^2} [1-\tilde{\pi}^{fin}(p)]          }    \Bigg]                               \eea
 The superscript $\textit{fin.}$ indicates the finite part of $\tilde{\pi}(p)$. \\
\indent Considering the perturbative nature of $\tilde{\pi}^{fin.}(p)$, $\delta Z_{\mathcal{B}}$ and taking into account the hierarchy of scales $m^2\gg M^2$, a useful approximation can be derived
\bea  m^2_+= M^2(1-\delta Z_\mathcal{B})\ , \ m^2_-(p)=m^2(1+\tilde{\pi}^{fin.}(p)) (1+\delta Z_\mathcal{B})             \eea
\indent The renormalized propagator can be written as
\bea \mathcal{P}_{\mu \nu \alpha \lambda}^R(p)=\frac{m^2}{  (1-\tilde{\pi}^{fin.}(p)) }\Bigg[\frac{iP^{[1e]}_{\mu\nu,\alpha\lambda}}{\big(p^2-m^2_+\big)\big(p^2-m^2_-(p)\big)}\Bigg]\label{prop1}\eea 

\indent Then, near the $m^2_+$ pole, it acquires the following form
\bea \mathcal{P}_{\mu \nu \alpha \lambda}^R(p)=\frac{m^2}{  [1-\tilde{\pi}^{fin.}(m^2_+)] }\Bigg[\frac{iP^{[1e]}_{\mu\nu,\alpha\lambda}}{\big(p^2-m^2_+\big)\big(m^2_+-m^2_-(m^2_+)\big)}\Bigg]\label{prop2}\eea 

\indent Since $m^2_+<m^2_{-}(m^2_+)$, the normalization is indeed positive and the unit residue can be obtained by a suitable fixation of $\delta Z_{\mathcal{B}}$. \\
\indent Accordingly, choosing a counterterm parameter $m^2$ such that $m^2>4m^2_e$, the self-energy $\pi^{fin.}(p^2)$ evaluated at the extra pole $\tilde{m}^2_p\approx m^2$ contains an imaginary part. This extra pole explicitly defined below
\bea \tilde{m}^2_p=m^2(1+\mathcal{R}[\tilde{\pi}^{fin.}(\tilde{m}^2_p)]  )(1+\delta Z_\mathcal{B})   \eea
is shifted as
\bea \tilde{m}^2_p+i\gamma                       \eea
with $\gamma=m^2\Im[\tilde{\pi}^{fin.}(\tilde{m}^2_p)](1+\delta Z_\mathcal{B})>0$. The  $\mathcal{R} $ and  $\mathcal{I}$ operators take the real and imaginary parts of a function, respectively. \\
\indent Since the associated residue is negative and $\gamma>0$, this extra excitation becomes a Merlin mode \cite{merlin}. This is a special kind of resonance. Being unstable, it does not  contribute to the asymptotic spectrum, playing no role in the unitarity constraints of the theory.

\subsection{Smooth massless limit for the renormalized saturated amplitude}

\indent After establishing a convergent path integral with smooth massless limit, thanks to the presence of the (KON) field, it is possible to evaluate the saturated amplitude for the renormalized physical asymptotic mode \cite{gab}
\bea \mathcal{A}(k)=-\frac{1}{2}T^{*\mu \nu}(k)\mathcal{P}_{\mu \nu \alpha \lambda}^R(k)T^{\alpha \lambda}(k)=\frac{i}{2}T^{*\mu \nu}(k)\frac{P^{[1e]}_{\mu\nu,\alpha\lambda}(k)}{\big(k^2-m^2_+\big)}T^{\alpha \lambda}(k)\eea
\indent The general structure of the c-number sources is constrained by the symmetries of the kinetic operator
\bea T_{\mu \nu}(k)=\big(k_\mu Y_\nu(k)-k_\nu Y_\mu(k)\big) \eea
with, $Y_\mu(k)$ being a general vector. This structure is in accordance with the conditions imposed by the auxiliary fields. \\
\indent The amplitude explicitly reads
\bea \mathcal{A}(k)= iY^{*\mu}(k)\frac{k^2\theta_{\mu \nu}}{(k^2-m^2_+)}Y^\nu(k)\eea
\indent Then, considering the frame $k_\mu=(m_+,0,0,0)$ for the massive case, and the chosen metric signature, the associated residue indeed has an imaginary part whose sign is in compliance with unitarity.\\
\indent On the other hand, taking the massless limit yields 
\bea \mathcal{A}(k)=i \frac{k^2|Y_\mu(k)|^2-|k_\alpha Y^\alpha|^2}{k^2}\eea
whose pole lies in the spin-$0$ sector associated with the longitudinal part of the sources. This is also a unitary pole.

\section{Considering the master action approach}\label{sec7}

\indent It is worth mentioning that the complete master action approach also demands the inclusion of a current squared term for the Kalb-Ramond Lagrangian  in order to establish the weak duality relations \cite{spa,gab} between
\bea S=\int d^4x\Big[-(\p^\mu \mathcal{B}_{\mu \nu})^2+\frac{M^2}{2}\mathcal{B}_{\mu \nu}\mathcal{B}^{\mu \nu}+\frac{\sqrt{2}}{M}\p^\mu \mathcal{B}_{\mu \nu}J^\nu-\frac{J_\mu J^\mu}{2M^2}      \Big] \eea
and the Proca model
\bea S=\int d^4x\Big[-\frac{1}{4}\mathcal{F}_{\mu \nu}\mathcal{F}^{\mu \nu}+\frac{M^2}{2}\mathcal{A}_\mu\mathcal{A}^\mu-\mathcal{A}_\mu J^\mu  \Big]                                             \eea
defined up to the auxiliary fields. The c-number current $J_\mu(x)$ is added to achieve a map between the functional generators.
We use the definition for the curvature $\mathcal{F}_{\mu \nu}=\p_\mu \mathcal{A}_\nu-\p_\nu \mathcal{A}_\mu$. This master action formulation can be immediately generalized for any $D\geq 2+1$ dimensions. In order to establish this specific master action duality, the mass parameter associated to the coupling should be the same as the one defining the dispersion relation. \\
\indent The quadratic terms on the sources are the origin of the so-called contact terms appearing in the dual map. Here, we extend it for the case of operator valued spinor sources $J_\mu(x)=e\Bar \Psi(x) \gamma_\mu \Psi(x)$. Then,  it is possible to consider a Hubbard-Stratonovich transformation
\bea -\frac{J_\mu J^\mu}{2M^2}=\frac{M^2}{2}\Omega_\mu \Omega^\mu-\Omega_\mu J^\mu\eea
by means of the auxiliary field $\Omega_\mu(x)$ being implicitly assumed that $\mathcal{D}\Omega_\mu$ is also introduced in the measure.\\
\indent Then, the complete additional action terms, beyond the auxiliary ones, read
\bea \mathcal{L}'=\frac{\sqrt{2}}{M}\p^\mu\mathcal{B}_{\mu \nu} \big(e\bar{\Psi} \gamma^\nu \Psi\big) +\Bar{\Psi}\Big(i\slashed{\partial}-m_e\Big) \Psi +\frac{M^2}{2}\Omega_\mu \Omega^\mu-\Omega_\mu e\Bar \Psi(x) \gamma^\mu \Psi(x)        \eea

\indent There is still a question on renormalizability. One cannot take it for granted, considering the complicated tensor structure of the bosonic two-point part. The form of the self-energy tensor  as well as its derivative order are equally important in this respect. Then, the quadratic  piece of the bosonic quantum action  reads \footnote{For a quantum field generically denoted as $\mathcal{Q}_A(x)$, we consider its Fourier transform as $\mathcal{Q}_A(x)=\int \frac{d^4p}{(2\pi)^4}\mathcal{Q}_A(p)e^{-ip.x}$.}
\begin{align} \Gamma^R[\mathcal{B},\Omega]=&\int \frac{d^4p}{(2\pi)^4}\Bigg\{\ \frac{\mathcal{B}^{\mu \nu\ }(p)}{2}\Bigg[   P^{[1b]}M^2-P^{[1e]}\Big(p^2-M^2+\frac{2p^4}{M^2}\tilde \pi(p^2)\Big)\Bigg]_{\mu \nu \alpha \beta}\mathcal{B}^{\alpha \beta \ }(-p)\nonumber \\&   \frac{M^2}{2}\Omega_\mu(p) \Omega^\mu(-p)+\frac{1}{2}\Omega_{\mu}(p)p^2\theta^{\mu \nu}\tilde \pi(p^2)\Omega_\nu(-p) +\frac{\sqrt{2}}{M}\Omega_{\mu}(p)p^2\theta^{\mu \nu}\tilde \pi(p^2)p^\alpha \mathcal{B}_{\alpha \nu}(-p)\nonumber \\ &+\varepsilon^{\mu \nu \alpha \beta}\phi_\mu \p_\nu \mathcal{B}_{\alpha \beta}+\p^\mu \phi_\mu b\Bigg\}\label{e}     \end{align}

\indent As is already expected for a non-renormalizable theory, the on-shell renormalization conditions are not enough to eliminate the divergent pieces from the action. Then, extra counterterms that are not originally present in the Lagrangian are required to achieve this goal. This non-renormalizable content is kept even for the case in which the $\Omega_\mu$ field is absent, as in the previous section.\\
\indent As we are going to prove, the combination of the tensor and $\Omega_\mu$ field propagators imply that all purely fermionic 1(PI)  functions are the same as in the Proca theory with standard QED-like interaction. However, regarding the bosonic sector, the effect of the fermionic self-interaction associated with the dual map is taken into account after integrating the $\Omega_\mu(x)$ field. Its contributions start in order $e^4$.\\
\indent Interestingly, for the specific case of the fermion self-energy $\Sigma(p)$,  the Proca theory result is indeed reproduced at all orders of perturbation theory. It occurs due to the fact that the sum of the tensor field and the $\Omega_\mu$ bare propagators contracted with their respective vertex structures yields the bare Proca propagator in the internal line insertions
\bea
       \gamma^\mu\mathcal{G}_{\mu \nu}^{(0)\Omega}(p)\mathcal{O}\gamma^\nu+\Delta^{\mu \nu}_{\ \sigma \omega}p^\sigma \gamma^\omega \mathcal{O} \mathcal{P}^{(0)}_{\mu \nu \alpha \beta}(p)\Delta_{\tau \varepsilon}^{\alpha \beta}p^\tau \gamma^\varepsilon=\gamma^\mu \mathcal{O}P^{(0)phys.}_{\mu \nu}(p)\gamma^\nu \eea
with the first term denoting the auxiliary field bare propagator, while the last term represents the Proca one. The equality holds just for the physical transverse part. The function $\mathcal{O}$ denotes the loop contribution contracted with the boson propagator associated with this specific internal line. It is a matrix in spinor space depending on a set of loop and external momentum variables. Here, $p_\mu$ denotes the loop integration momentum flowing in this specific bosonic internal line. Therefore, considering this result, it is easy to conclude that all purely fermionic 1(PI) functions reproduce the ones from the Proca model.\\
\indent Although the theory described by means of the antisymmetric field is non-renormalizable, all the external bosonic lines attached to the complete $n$-point functions are associated with the combination
\bea        \Omega_\nu(x)+\frac{\sqrt{2}}{M}\p^\mu\mathcal{B}_{\mu \nu}(x)                                   \eea              
implying that, after the redefinition

\bea        \Omega_\nu(x)\to \Omega_\nu(x)- \frac{\sqrt{2}}{M}\p^\mu\mathcal{B}_{\mu \nu}(x)                                   \eea              followed by the integration of the antisymmetric  field as well as its auxiliary fields recovers the Proca model with all the $n$ point functions associated to the standard trilinear interaction. Therefore, the claim of \cite{god} is indeed correct, there is no full duality between Proca and Kalb-Ramond fields if the same kind of coupling is considered for both models. However, there is indeed an interpolating master action involving the inclusion of an extra fermionic self-interaction that is valid even if considering the complete set of quantum corrections.

\section{Including spin interactions for a confining potential} \label{sec8}

\indent The main idea of this section is to extend the approach of \cite{spa2} to derive an effective confining potential for QCD \cite{conf,conf1} in which a specific set of spin interactions is also taken into account. This improvement has  phenomenological and theoretical motivations \cite{conf3,conf4,conf5,conf6,conf7,conf8}.
 The interacting Lagrangian presents a well-defined local interaction between the Maxwell and Kalb-Ramond models, each one coupled with a given fermionic source. After integrating the antisymmetric and the vector fields, the resulting theory presents a Cornell-like confining static potential with the presence of spin interactions. These do not involve delta-like singularities. Moreover, these new terms are also relevant to the macroscopic phenomenology of the system. \\
\indent Beyond these semi-classical achievements, we are also interested in deriving a model in compliance with a set of good renormalization properties. Unlike the well-known Pauli-interaction case, this specific form of coupling does not imply the necessity of extra counterterms, not originally present in the bare action, to eliminate the divergences. However, these features are strongly dependent on the presence of suitable Lagrange multipliers restricting the path integral. The model is explicitly given below

\begin{align} \mathcal{L}&=-\frac{1}{4}\mathcal{F}^{\mu \nu}\mathcal{F}_{\mu \nu}+\frac{\alpha}{\sqrt{2}} M\mathcal{B}_{\mu \nu}\mathcal{F}^{\mu \nu}- \big(\partial^\mu\mathcal{B}_{\mu \nu}\big)^2+\frac{M^2}{2}\mathcal{B}_{\mu \nu}\mathcal{B}^{\mu \nu} +   \mathcal{A}_\mu J^\mu\nonumber\\ & +\p_\mu \mathcal{A}^\mu B+\phi_\sigma \epsilon^{\sigma\mu \nu \alpha}\p_\mu \mathcal{B}_{\nu \alpha} +b\p_\mu \phi^\mu+g\mathcal{B}_{\mu \nu}\sigma^{\mu \nu} +\Bar{\Psi}\Big(i\slashed{\p}-m_e  \Big)\Psi      \label{eqq}             \end{align}

with the same coupling structure described by a tensor field. $B$, $b$ and $\phi_\mu$ are Lagrange multiplier fields and $g$ is a coupling such that $g\ll e$. $\alpha$ is a dimensionless coupling approximately equal to the unit. The interaction tensors are defined as
 $J_\mu=e\Bar{\Psi}\gamma_\mu\Psi$ and $\sigma_{\mu \nu}=-\frac{i}{4}\Bar{\Psi}\Big[\gamma_\mu,\gamma_\nu\Big]\Psi$.\\
 \indent At this point, some clarifying remarks are demanded. Since there are no derivative couplings, it is easy to derive an expression for the superficial degree of divergence that does not increase with the number of vertices, a necessary condition for a renormalizable model. However, although the model is invariant under the well-known electromagnetic $U(1)$ symmetry, the coupling implies a non-linear breaking of the gauge symmetry associated with the Kalb-Ramond model. It implies that the quantum corrections may generate dynamics for both $P^{[1b]}$ and $P^{[1e]}$ sectors of the antisymmetric field, leading to a non-renormalizable model. However, the presence of the (KON) auxiliary fields guarantees a converging asymptotic path integral, even in the massless limit, and also eliminates radiative corrections for the non-dynamical sector. In order to address this last point, we first note that the $6$ off-shell degrees of freedom from the tensor source can be described  as
\bea \sigma_{\mu \nu}\equiv \epsilon_{\mu \nu \alpha \beta} \p^\alpha W^\beta+\p_\mu \tilde{J}_\nu-\p_\nu \tilde{J}_\mu   \eea
with the first term being associated to the non-physical projector. Here, both $W_\beta$ and $\tilde{J}_\beta$ are operator-valued vector fields. Note that the longitudinal parts do not contribute.\\
\indent Then, after the path integral redefinition with unit Jacobian
\bea \phi_\mu(x) \to \phi_\mu(x)+g\theta_{\mu \alpha}W^\alpha(x)                     \eea
the resulting coupling displays explicit independence from non-physical modes.\\
\indent Although the self-energy tensor for the Kalb-Ramond field has the following structure 
\bea \Sigma_{\mu \nu \alpha \beta}=P_{\mu \nu \alpha \beta}^{[1e]}\ \beta^{[1e]}(p^2)+P_{\mu \nu \alpha \beta}^{[1b]}\ \beta^{[1b]}(p^2)                                          \eea
just the first term contributes if the mentioned restriction on the functional generator is considered. More specifically, the delta function associated to the Lagrange multiplier  implies the following condition for all quantum paths
\bea \epsilon^{\mu \nu \alpha \beta}\p_\nu \mathcal{B}_{\alpha \beta}-\p^\mu b=0       \eea
which is compatible with the requirement $P_{\mu \nu \alpha \beta}^{[1b]}\mathcal{B}^{\alpha \beta}=0$. Moreover, since it is possible to prove that the self-energy correction has a divergent piece proportional to $p^2$, the two-point structure is renormalizable.\\
\indent Another equivalent description of such a scenario is given in terms of the (KON) approach. For the present interaction, the operator sum below
\bea \phi_\alpha(x)-gW_\alpha(x) \eea
is the one obeying the double pole free equation. Therefore, the definition of the physical subspace
\bea  \big(\phi_\alpha(x)-gW_\alpha(x)\big)^+|phys\rangle=0                 \eea 
is a function of the positive frequency part of the field combination above. Here, $|phys\rangle$ is a general designation for a physical state.\\
\indent The bosonic two-point part of the quantum action displays the effect of the quantum self-energies

\begin{align} \Gamma[\mathcal{B}]&=\int \frac{d^4p}{(2\pi)^4}\Bigg\{\ \frac{\mathcal{B}^{\mu \nu\ }(p)}{2}\Bigg[   P^{[1b]}M^2-P^{[1e]}\Big(p^2(1+\beta^{[1e]}(p^2))-M^2\Big)\Bigg]_{\mu \nu \alpha \beta}\mathcal{B}^{\alpha \beta \ }(-p)\nonumber \\ &
+\big(\frac{\alpha}{\sqrt{2}} M+\gamma(p)\big) \mathcal{F}^{\mu \nu}(p)\mathcal{B}_{\mu \nu}(-p)-\frac{1}{2}\mathcal{A}^\mu(p) p^2(1+\tilde{\pi}(p^2)) \theta_{\mu \nu}\mathcal{A}^\nu(-p)
\Bigg\}\end{align}
up to the auxiliary field Lagrangian. \\
\indent The function $\gamma(p)$ is the correction for the Gaussian coupling between the bosonic fields. Focusing on the $1$-loop contribution, it is related to a mixed graph involving both couplings with just fermionic internal lines. Therefore, considering the $U(1)$ invariance of the system, the loop integration symmetry as well as the fact that the trace of the product of an odd number of gamma matrices vanishes, one concludes that the divergent part of this radiative correction is constant.  Since it can be absorbed in a renormalization procedure associated with the parameter $\alpha$, the mixed two-point part is indeed renormalizable. However, a complete analysis of the whole system demands the investigation of the higher $n$-point functions.\\
\indent As a semi-classical approximation, one replaces the operator-valued spinor sources by their vacuum averages. In this manner, considering the expression derived in \cite{spind}, in the ultra non-relativistic limit, results in 

\bea J_\mu=e\delta_0^\mu\Big[\delta^3(\vec x-\vec x_1)-\delta^3(\vec x-\vec x_2)                    \Big]\eea

\bea   \sigma_{\mu \nu}=-\frac{1}{2}\Delta_{\mu \nu}^{\ ij}\epsilon_{ijk}\big(\langle \sigma^k_1 \rangle \delta^3(\vec x  -\vec x_1)+\langle \sigma^k_2 \rangle \delta^3(\vec x  -\vec x_2)\big)     \eea                                                          
with the particles, possessing charges and spin, being placed at $\vec x_1$ and $\vec x_2$ spatial positions. Here, the  spin of a particle $I=1,2$ is given by $S^k_I=\frac{1}{2}\langle \sigma^k_I \rangle $ in natural units. The average is taken with the asymptotic electron spinor. The operators $ \sigma^k$ are the Pauli matrices. \\
\indent Then, integrating out all the bosonic fields from eq. (\ref{eqq}) yields
\bea \mathcal{L}=-\frac{1}{2}\frac{J_\mu(\Box+M^2) J^\mu}{\Box^2}-\frac{Mg}{\sqrt{2}}\frac{J^\mu \p^\alpha \sigma_{\alpha \mu}}{\Box^2}  +\frac{(\p^\alpha \sigma_{\alpha \mu})^2g^2 }{\Box (\Box+M^2)}\Big(1-\frac{M^2}{4\Box}      \Big)                                 \eea
up to the fermionic kinetic operators. We considered $\alpha \approx 1$, in accordance with the latter developments. This is clearly just the first approximation for the effective  Lagrangian, since no self-energy correction due to the integration of fermionic quantum fluctuations is considered.\\
\indent The interparticle potential can be inferred from the effective Lagrangian leading to
\begin{align} V(r)=& -\frac{e^2}{4\pi r}+\frac{e^2M^2r}{8\pi}-\frac{g^2}{32\pi r}\langle \vec \sigma_1\rangle .\langle \vec \sigma_2\rangle\Big(1-5e^{-Mr}  \Big)   -\frac{g^2}{64\pi}\langle \sigma_1^m\rangle .\langle \sigma_2^n\rangle \Big(\frac{\delta_{mn}r^2-r_mr_n}{r^3}  \Big) \nonumber \\&-\frac{5g^2}{32\pi M^2}\langle \sigma_1^m\rangle .\langle \sigma_2^n\rangle\Bigg[ \Big(\frac{-r^2\delta_{mn}+3r_mr_n }{r^5}\Big)\Big(1-e^{-Mr}(Mr+1)\Big)-\frac{r_mr_nM^2e^{-Mr}}{r^3} \Bigg]
\end{align}
with  $r\equiv |\vec x_1-\vec x_2|$.\\ 
\indent The approximated expression for the potential related to the electric charge has a Coulomb part and a linear confining contribution, important for the long-distance behavior. The spin interactions have no delta-like singularities and are also relevant for the macroscopic phenomenology of the system. Their contribution is subleading due to the smallness of the coupling $g$. Analogously to the QED case, the self energy corrections induce modifications in the potential, possibly implying in delta-like contributions that may produce shifts in the energy spectrum. This is an interesting future perspective for these last developments.

\section{Conclusions and Perspectives}\label{sec9}

\indent This article provided several discussions on the properties of the Kalb-Ramond model. Namely, we extended the so-called (KON) formalism to describe the complicated auxiliary sector of this theory. Two extra fields were introduced in order to define a second-class system and proceed with the quantization without any ambiguities, considering the constraints and equations of motion in their strong forms. It leads to a covariant description of the system suitable to discuss the massless limit and the associated jump in the degrees of freedom. It also furnished a well-defined framework to establish the weak dual correspondence of the massive/massless theory with the Proca/scalar field models.\\
\indent The $D$ dimensional generalization of the Stueckelberg procedure was performed to derive a well-defined massive/massless phase transition. It is compatible with the previous developments, since the (KON) formalism is in accordance with the path integral description. It allowed the derivation of Ward-Takahashi identities, even for the massive theory. A set of interactions were also investigated, since they clearly define the limits of the dual correspondence. In this way, even the weak dual correspondence between Proca and  the massive Kalb-Ramond model is broken due to radiative corrections for the interacting system. The emergence of undesired extra modes due to quantum corrections is consistently avoided by a well-defined renormalization procedure. These excitations become unstable and do not contribute to the asymptotic spectrum. The analysis of the saturated amplitude revealed an interesting discussion on the degree of freedom variation in the massless limit of the renormalized system.\\
\indent Considering the master action approach, regarding the fermionic sector, all the   1(PI) functions from the Proca electrodynamics are recovered by the dual Lagrangian. On the other hand, the theory described by the tensor field is non-renormalizable and also presents a set of physical differences from the Proca model.\\ 
\indent The final section was  devoted to extending a previously discussed confining potential \cite{spa2}, involving the Kalb-Ramond field, to include spin interactions. We embedded the model in a renormalizable structure thanks to the presence of auxiliary fields. The spin interaction is free from delta-like singularities and is relevant even for macroscopic distances. The idea was to improve the effective potential, taking into account a wider variety of aspects related to QCD low-energy phenomenology. \\
\indent Regarding the future perspectives, the present research led to some complementary ideas. Although the Stueckelberg procedure furnished a degree of freedom conserving structure to the model, the tensor field sector indeed presents a jump in its degrees of freedom at $m \to 0$. The interesting point is the possibility to use it as a phase transition mechanism. More specifically, the quantum perfect fluid description associated with the massless model can be understood as a phase with some frozen degrees of freedom, a limiting case of the massive one. Then, some questions may arise. Which is the system represented by the massive model, is it a generalization of the superfluid description \cite{fluid2} mentioned in this article? Which are the set of physical implications? Moreover, regarding interactions, a weak dual description by means of a tensor field provides a wider set of possible couplings, improving the modeling power. It also motivates a more detailed discussion on the nature and  field theory status of such new interactions. From these considerations, new achievements from foundations to particle physics can be obtained. The possibility of an interacting Kalb-Ramond plasma analyzed in a finite temperature framework can be a source of numerous discussions involving the massless transition. Accordingly,  regarding the final section on spin-dependent interactions, at least the two-point part of the quantum action is renormalizable. However, a complete investigation with the explicit form of the quantum corrections as well as the consideration of vertex and higher $n$-point functions is still missing. This is another correlated future perspective. Moreover, the idea of including screening effects from radiative corrections for the confining interaction potential is a feasible goal that resembles the well-known Lamb shift discussion in the context of standard electrodynamics.

\section{Appendix A: distributions}
\indent The so-called Pauli-Jordan distribution is defined by its initial conditions and associated differential equations. It can be decomposed into positive/negative frequency parts. These features are displayed below
\begin{align}&i\Delta(x-y, s)=\Delta^+(x-y, s)+\Delta^-(x-y, s) , \nonumber\\  &  \tilde \Box \Delta(x-y,s)=-s\Delta(x-y,s),\quad  \Delta(x-y,s)|_{x_0=y_0}=0 ,\nonumber\\  &  \Delta^\pm(x-y, s)=\mp \frac{1}{(2\pi)^2}\int d^3p\ \delta(\tilde p^2-s)\Theta(\pm p_0)e^{-ip.(x-y)} ,\nonumber\\  &  \quad \p_0\Delta(x-y,s)|_{x_0=y_0}=-\delta^2(x-y)
\end{align}
\indent Analogously to the Pauli-Jordan function, the so-called double pole distribution is defined below
\begin{align} &(\tilde \Box+s)E(x-y,s)=\Delta(x-y,s)  \ , \ \p_0^3E(x-y,s)|_{x_0=y_0}=-\delta^2(x-y)\ , \nonumber \\
&E(x-y,s)|_{x_0=y_0}=0\ ,\ 
E(x,m^2_I)\equiv-\int d^3u\ \varepsilon(x,0,u)\Delta(x-u,m^2_I)\Delta(u,m^2_I)  \end{align}
\indent It also has positive and negative frequency parts $ iE(x-y,s)=E^+(x-y,s)+E^-(x-y,s)$, defined as
 \bea E^\pm(x-y,s)=\pm (2\pi)^{-2}\int d^3p \theta(\pm p_0)\frac{d\delta(\tilde p^2-s^2)}{d\tilde p^2}e^{-ipx}\eea

\section{Appendix B: Spin projectors}

Using the spin-0 and spin-1 projection
operators acting on vector fields, respectively,

\bea  \omega_{\mu\nu} = \frac{\p_{\mu}\p_{\nu}}{\Box} \ , \
\theta_{\mu\nu} = \eta_{\mu\nu} -
\frac{\p_{\mu}\p_{\nu}}{\Box}\quad , \label{pvectors} \eea
as building blocks, one can define the projection
operators in $D$ dimensions acting on antisymmetric rank-2
tensors.\\
\indent The above set of operators satisfies 
\begin{eqnarray}
\theta^2 = \theta , \quad \omega^2 = \omega \quad \textmd{and} \quad \theta \omega = \omega \theta = 0. 
\end{eqnarray}

On the other hand, the set of the antisymmetric four-dimensional Barnes-Rivers operator are given by  
\begin{equation}
P^{[1b]}_{\mu\nu,\alpha\lambda} = \frac{1}{2}(\theta_{\mu\alpha} \theta_{\nu\lambda} - \theta_{\mu\lambda} \theta_{\nu\alpha}),
\end{equation}
\begin{equation}
P^{[1e]}_{\mu\nu,\alpha\lambda} = \frac{1}{2}(\theta_{\mu\alpha} \omega_{\nu\lambda} + \theta_{\nu\lambda} \omega_{\mu\alpha} - \theta_{\mu\lambda} \omega_{\nu\alpha} - \theta_{\nu\alpha} \omega_{\mu\lambda}).
\end{equation}
They satisfy the very simple algebra
\begin{eqnarray}
&(P^{[1b]})^2 = P^{[1b]}, \quad (P^{[1e]})^2 = P^{[1e]},&\nonumber\\
&P^{[1b]}P^{[1e]} = P^{[1e]}P^{[1b]}=0.
\end{eqnarray}

and the completeness relation
\bea  {P^{[1e]}}_{\mu\nu}^{\ \alpha\lambda}+{P^{[1b]}}_{\mu\nu}^{\ \alpha\lambda}=\frac{1}{2}\big( \delta_\mu^\alpha \delta_\nu^\lambda -\delta_\mu^\lambda \delta_\nu^\alpha      \big)             \eea

\acknowledgments
\indent G.B. de Gracia thanks the São Paulo Research Foundation -- FAPESP grant No. 2021/12126-5. He also thanks Dr. R. da Rocha and Dr. B.M. Pimentel for useful discussions.

\end{document}